\documentclass[12pt]{article}

\usepackage{scicite}
\usepackage{times}
\usepackage{tabu}
\usepackage{graphicx}
\usepackage{multirow}
\usepackage{url}

\newcommand{\etal}{\textit{et al.}}
\newcommand{\gomez}{G\'{o}mez }

\newenvironment{sciabstract}{%
\begin{quote} \bf}
{\end{quote}}

\title{Geodesy of irregular small bodies via neural density fields: geodesyNets}

\author
{Dario Izzo$^{1\dagger\ast}$ and Pablo \gomez$^{1\ast}$\\
\\
\normalsize{$^{1}$Advanced Concepts Team, European Space Agency}\\
\normalsize{European Space Research and Technology Centre (ESTEC)} \\
\normalsize{Keplerlaan 1, 2201 AZ Noordwijk, The Netherlands}\\
\\
\normalsize{$^\dagger$To whom correspondence should be addressed; E-mail:  dario.izzo@esa.int .} \\
\normalsize{$^\ast$These authors contributed equally to this work.}
}

\date{}

\begin{document} 


\baselineskip24pt


\maketitle 

\hyphenation{Chur-yu-mov Gera-si-men-ko Chur-yu-mov-Gera-si-men-ko}


\begin{sciabstract}
    Asteroids' and comets' geodesy is of increasing interest in a wide range of fields ranging from astronomy to potential future mining prospects, as well as being essential to successful proximity operations of spacecraft. However, the problem of inferring the internal density distribution of irregular bodies from gravity measurements is challenging and ill posed, with the added difficulty of finding compact representations of the gravitational field surrounding them. 
    We present a novel approach based on artificial neural networks, so-called geodesyNets, and present compelling evidence of their ability to serve as accurate geodetic models of highly irregular bodies using minimal prior information on the body. The approach does not rely on the body shape information but, if available, can harness it. GeodesyNets learn a three-dimensional, differentiable, function representing the body density, which we call neural density field. The body shape, as well as other geodetic properties, can easily be recovered. We investigate six different shapes including the bodies 101955 Bennu, 67P Churyumov–Gerasimenko, 433 Eros and 25143 Itokawa for which shape models developed during close proximity surveys are available. Both heterogeneous and homogeneous mass distributions are considered. The gravitational acceleration computed from the trained geodesyNets models, as well as the inferred body shape, show great accuracy in all cases with a relative error on the predicted acceleration smaller than 1\% even close to the asteroid surface. When the body shape information is available, geodesyNets can seamlessly exploit it and be trained to represent a high-fidelity neural density field able to give insights into the internal structure of the body.
    This work introduces a new unexplored approach to geodesy, adding a powerful tool to consolidated ones based on spherical harmonics, mascon models and polyhedral gravity. Compared to previous approaches, geodesyNets are generic, they maintain accuracy and computational efficiency and do not rely on the knowledge of the body shape model.
\end{sciabstract}

\section*{Introduction}
The last two decades have witnessed the beginning of space exploration aimed at minor solar system bodies, such as asteroids and comets, beyond simpler fly-by missions.
The mission NEAR visited the asteroid 253 Mathilde and was later successfully inserted into an initial orbit around 433 Eros in February 2000. From there it thoroughly studied the asteroid and eventually performed a successful touch down a year later as the mission ended \cite{veverka2000near}. 
After that, in 2005, the sample return mission Hayabusa, briefly touched down on the Muses Sea of 25143 Itokawa to collect samples from the asteroid surface and bring them back to Earth \cite{yano2006touchdown, fujiwara2006rubble}. 
In 2011 the spacecraft Dawn surveyed Vesta for 12 months to then leave and reach, in 2015, its final destination, Ceres. 
In 2014, the Rosetta spacecraft and its Philae lander were able to visit and land on 67P Churyumov-Gerasimenko examining at close proximity the activity of the frozen comet as it approached the Sun \cite{capaccioni2015organic}.
Hayabusa 2 sampled 162173 Ryugu \cite{watanabe2019hayabusa2} in 2018 and later returned the sample to Earth. Most recently, in 2020, OSIRIS-Rex obtained a sample from 101955 Bennu \cite{lauretta2017osiris}. 
The continuation of this trend is clear as several missions are planned for this decade such as Hera \cite{michel2018european}, ZhengHe \cite{jin2020simulated} and Psyche \cite{lord2017psyche}.
Sampling and surveying asteroids and comets provide unique opportunities to study the history and development of the solar system \cite{glassmeier2007rosetta, connolly2016chondrules, herbst2021macroporosity}. 
Interest in visiting and surveying small solar system bodies was exclusively scientific up to recently, when several commercial entities showed interest into prospective asteroid mining and - with human space flight ambitions once again looking beyond Earth orbit - since topics surrounding in-situ resource utilization on minor planets are now of particular interest \cite{hein2020techno, calla2018asteroid, ross2001near}. 

In these types of interplanetary missions, knowledge of the geodesy of the investigated bodies plays a critical role in successfully performing orbital and surface proximity operations, in closely tracking touch-and-go trajectories as well as in evaluating the collected measurements and observations. The gravity field generated by the body and the acceleration induced on the spacecraft allow the precise planning and execution of mission operations \cite{hashimoto2010vision, accomazzo2017final}, while knowledge of the body shape and of its internal mass distribution - which may give insight into the body's origin and composition - are of interest to both scientists and mission operators \cite{kanamaru2019density, scheeres2020heterogeneous, braun2018geophysical}. 
The gravitational field of a celestial body is, for most operational purposes, typically represented by a spherical harmonics expansion of the gravitational potential with coefficients learned via Kalman filtering techniques \cite{hashimoto2010vision, accomazzo2017final}. Unfortunately this approach loses its appeal as the body irregularities become more important \cite{hirt2017convergence, sebera2016spheroidal}. Other options such as mascon models \cite{werner1996exterior, wittick2019mixed} and polyhedral gravity \cite{paul1974gravity, d2014analytical} can overcome some of these difficulties, but also introduce other requirements such as the need for a precise shape model or the assumption of a homogeneous internal density.

In this work, we introduce geodesyNets, a new, generic and unified approach to gravity representation which allows inversion even in those cases where other representations and techniques fail.
For geodesyNets, we take inspiration from recent trends and breakthroughs in artificial intelligence and computer vision, such the Generative Query Networks by Eslami \etal{} \cite{eslami2018neural} or the Neural Radiance Fields by Mildenhall \etal{} \cite{mildenhall2020nerf, park2020deformable, schwarz2020graf} who introduced novel neural network architectures and training methods for three-dimensional scene reconstruction from two-dimensional images. 
In a more abstract sense, their works follow a general, emerging trend of utilizing neural networks as a method for solving inverse problems \cite{padmanabha2021solving,xu2019neural,kim2020inverse,gomez2018laryngeal}. 
With geodesyNets, we solve the body-specific problem of gravity inversion and shape reconstruction using a neural network as approximator for the distribution of mass in an enclosing cubic volume. The investigated bodies in our test cases are asteroids or comets. 
We are able to reach competitive accuracy compared to previous approaches with notably fewer assumptions about body shape and density distribution. The trained geodesyNets are able to simultaneously represent both, body shape and density distribution, accurately. Further, the approach is able to incorporate body shape information for improved results. With shape information, we demonstrate the successful application to bodies with heterogeneous density distributions. Finally, we show that training geodesyNets is a computationally efficient process, which bears the promise of on-board applicability. To allow for an easy replication of our results, we provide all code\footnote{\url{https://github.com/darioizzo/geodesynets}} and data online\footnote{\url{https://zenodo.org/record/4749715#.YJrR6OhfiUk}}.

\section*{Results}

\subsection*{Geodesy Artificial Neural Networks: geodesyNets}
We represent and study the geodetic properties of generic celestial bodies using fully connected neural networks that we call geodesyNets. 
A geodesyNet represents the body density directly as a function of Cartesian coordinates. 
Recently, \emph{Mildenhall et al.} \cite{mildenhall2020nerf} introduced a similar architecture called Neural Radiance Fields (NeRF) to represent three-dimensional objects and used it to reconstruct complex scenes with an impressive accuracy learning from a set of two-dimensional images. 
The training of a NeRF solves the inverse problem of image rendering as it back-propagates the difference between images rendered from the network and a sparse set of observed images.
Similarly, the training of a geodesyNet solves the gravity inversion problem. The network learns from a dataset of measured gravitational accelerations back-propagating the difference to the corresponding accelerations computed from the density represented by the network.
\begin{figure}[ht]
    \centering
    \includegraphics[width=0.8\columnwidth]{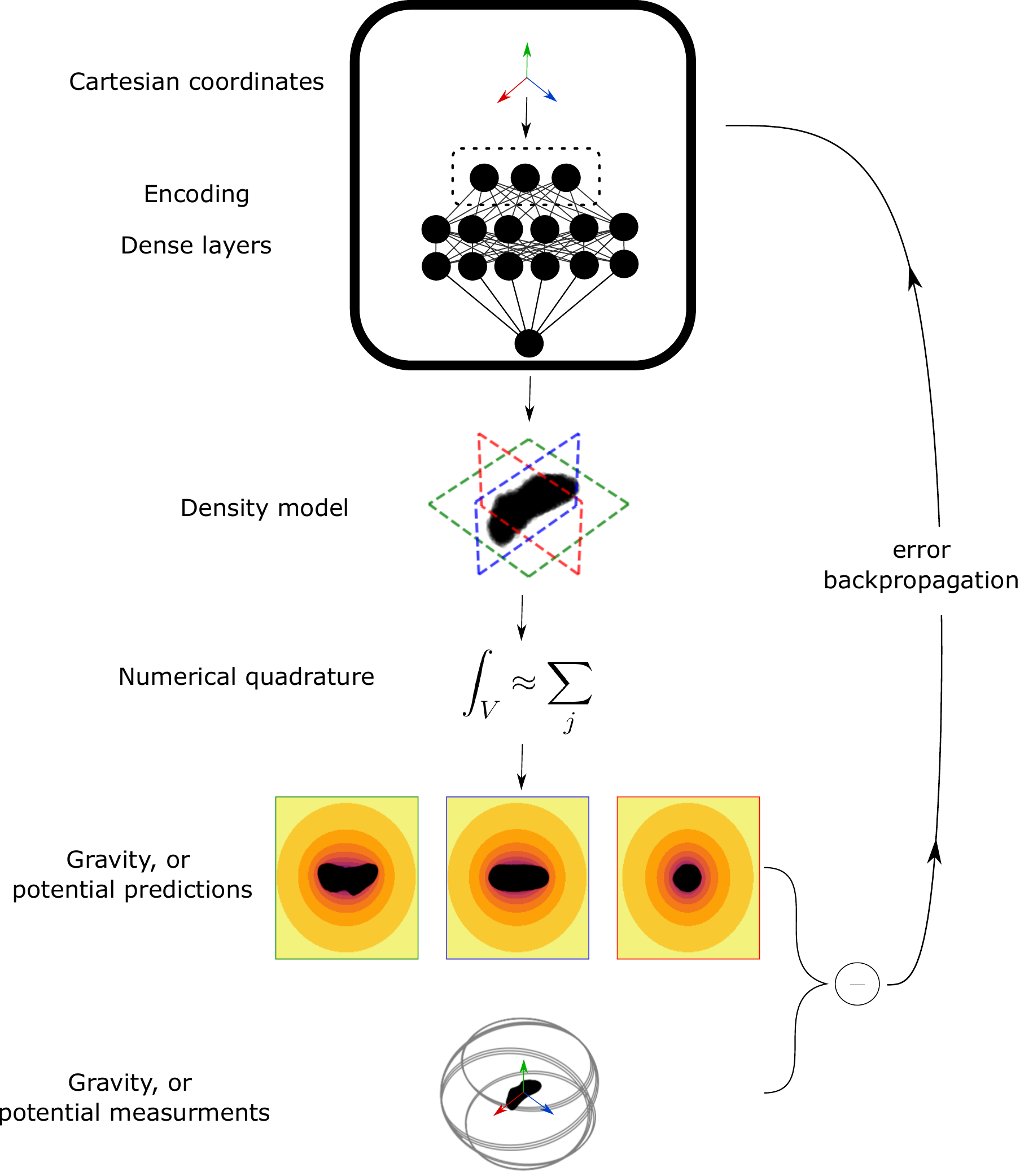}
    \caption{Overall schematics of the process of training a geodesyNet. }
    \label{scheme}
\end{figure}
At the end of its training, the network provides a differentiable expression mapping the position within a cubic volume $V$ to a candidate body density $\rho(x,y,z)$ compatible with the observed accelerations. Analogously to the radiance fields of the NeRF, we refer to $\rho$ as a neural density field. The neural density field can be used to study the body's internal structure, to compute the gravitational potential field and accelerations outside and inside the body, as well as to derive quantities such as the spherical harmonics coefficients that depend uniquely on said density function.
Because of the mathematical properties of artificial neural networks as universal approximators, recently re-discussed in depth by \emph{Calin} \cite{calin2020universal}, sharp density discontinuities and sudden variations of the body structure can also be represented and, in fact, a geodesyNet implicitly learns a plausible representation for the asteroid surface as a two-dimensional discontinuity embedded in the three-dimensional space. In other words, indicating with $V_{B}$ the volume actually occupied by the body, a geodesyNet learns to infer a vanishing density outside $V_{B}$ and to jump to a finite value when approaching the body surface $\partial V_{B}$, without ever being trained on where $\partial V_{B}$ actually is.

The overall architecture of a geodesyNet is shown in Figure 1 (more details are given in the method section). 
First, the Cartesian coordinates $x,y,z$ -- indicating the position of a point within the hypercube $V$ -- are fed into an encoding layer mapping them to a representation suitable for the network. 
\emph{Rahaman et al.} \cite{rahaman2019spectral} have recently noted how the expressivity property of deep networks, able to fit even random input-output mappings, comes with a spectral bias manifesting in the network learning low-frequencies first. This is important to the geodetic application proposed here as an irregular body shape, to start with, might suffer from a poor representation of important high frequency contents, for example at the surface $\partial V_{B}$. The encoding layer allows experimenting with transformations of the Cartesian coordinates able to control the network's spectral bias. Eventually, we find that a direct Cartesian encoding, coupled to periodic activation functions between layers \cite{sitzmann2020implicit}, offers optimal performance in terms of the resulting quality of the neural density field (details in the supplementary material Table S5).
After the encoding layer, a number of fully connected layers follow, forming the main body of the network with its learnable parameters.
The network output, i.e. the neural density field, is then integrated numerically over the volume $V$. The numerical quadrature necessitates evaluation of the network at $N$ distinct points inside the hypercube $V$. The quadrature is used to derive the gravitational acceleration denoted generically with $\hat y_i$ in $M$ points where measurements $y_i$ are available. The difference between the computed and the measured (ground-truth) values is fed into a loss function $\mathcal L(\theta)$ which is used to backpropagate the error and, thus, to update the model parameters $\theta$.

\begin{figure}[ht]
    \centering
    \includegraphics[width=0.8\columnwidth]{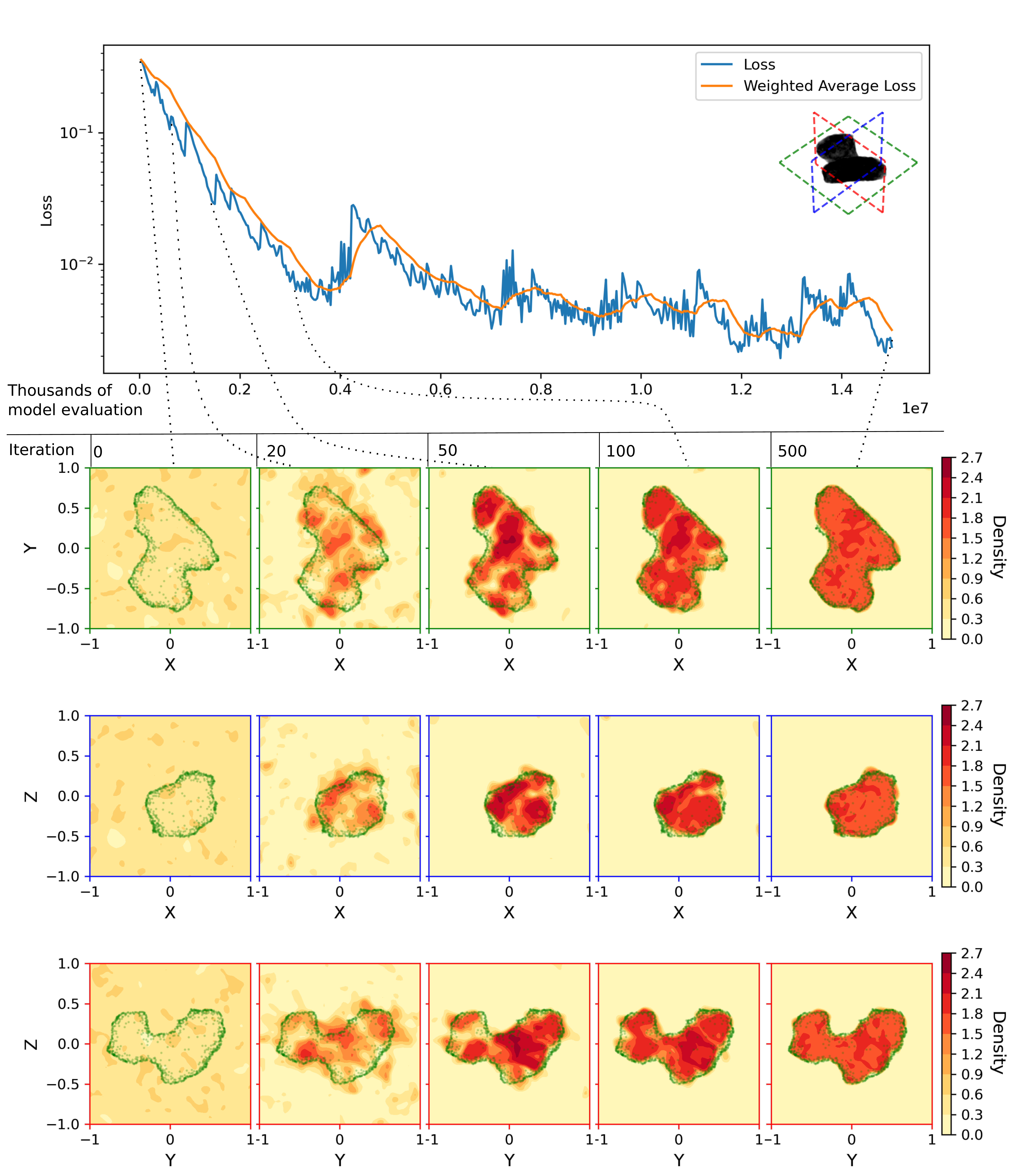}
    \caption{Exemplary training of a geodesyNet on the gravitational acceleration of a homogeneous density Churyumov-Gerasimenko comet model. Non-dimensional units are used (see details in the Methods section). The training loss is shown on the top while the learned mass distribution in the hypercube $V = [-1,1]^3$ is show in the bottom for the $xy$, $xz$ and $yz$ cross sections. As learning progresses, the mass distribution evolves to reconstruct the correct asteroid shape and a uniform internal density.}
\end{figure}

The result is a process that gradually learns a three-dimensional model of the body density compatible with the measured gravitational field, as seen, for example, in Figure 2, where the learning process is shown in the case of synthetic gravitational data generated for the comet Churyumov-Gerasimenko. The learned model can be used to determine the geometric shape and (to some extent) the internal structure of the body, its orientation in space, and its gravity field, thus allowing the geodetic properties of the body to be fully determined from the parameters $\theta$ defining the neural architecture. The same pipeline can also be applied to measurements of the gravitational potential, resulting in a neural density field with similar accuracy.

\begin{figure}[ht]
    \centering
    \includegraphics[width=0.8\columnwidth]{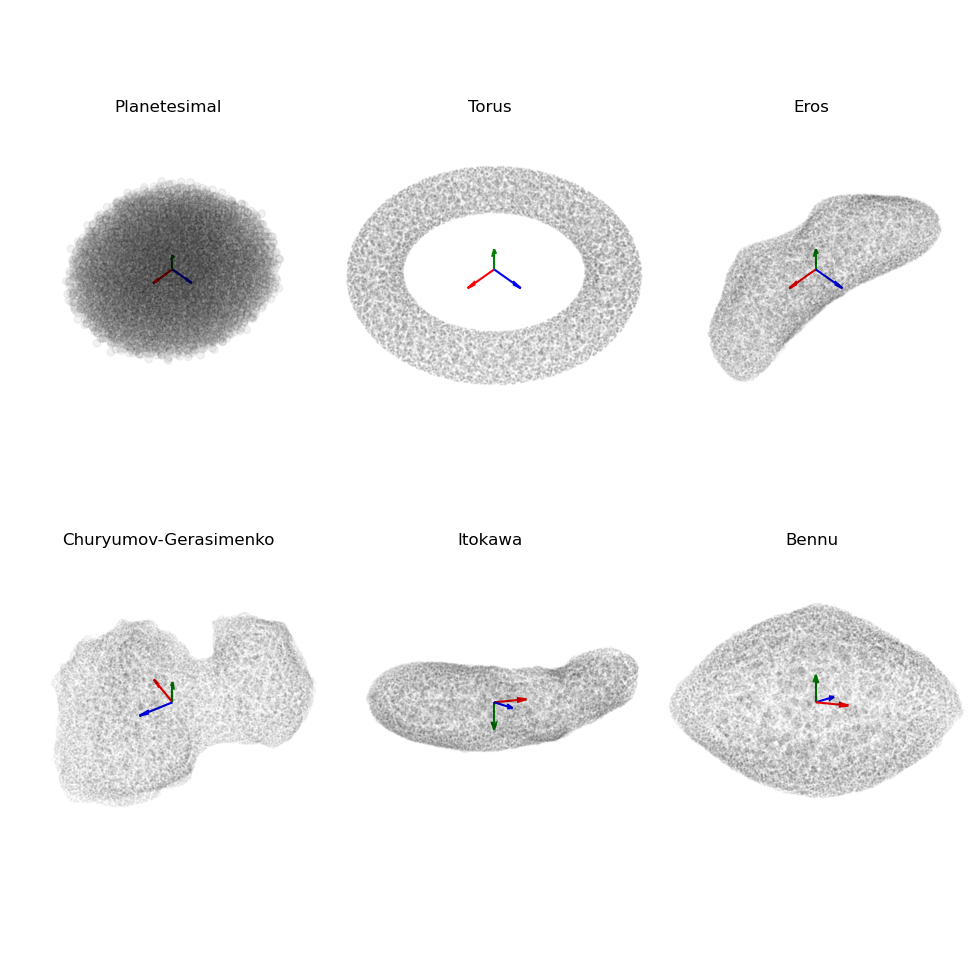}
    \caption{Mascon models for the six bodies considered. The dimension of each mascon is constant, hence the actual mass distribution inside the bodies is not visible, only their shape. Principal axes, used throughout this work, are also shown for convenience.}
\end{figure}

\subsection*{The ground truths}
To produce synthetic values for the measurements $y_i$ of gravitational accelerations, we use mascon models representing mass distributions with varying degrees of heterogeneity. Each mascon model is a list of tuples $\mathcal M = \left\{(x_i, y_i, z_i, m_i)\quad i=1..n\right\}$ and allows to compute the ground-truth gravitational acceleration at a generic point $\mathbf r_i$ via the formula:
$$
\mathbf a(\mathbf r_i)  = - G \sum_{j=1}^{n} \frac {m_j}{r_{ij}^3}\mathbf r_{ij} \, ,
$$
where $G$ is the Cavendish constant, and $m_j$ is the generic mascon mass placed at $\mathbf r_j$. 
We consider the asteroids 433 Eros, 25143 Itokawa and 101955 Bennu and the comet 67P Churyumov–Gerasimenko, as well as a fictitious Planetesimal and a toroidal shaped body we call Torus. First, we generate a mascon model  for each body that represents a homogeneous mass distribution. Then, in the case of Bennu, Itokawa and Planetesimal we generate additional mascon models representing a heterogeneous mass distribution. For Bennu, we introduce an equatorial region with lower density with a reduction factor $f=2$. For Itokawa we make the asteroid head heavier augmenting its density by a factor $f=1.6$. For Planetesimal, instead, we create an internal cavity of spherical shape, thus radically changing the body topology.

The resulting mascon models are visualized in Figure~(3). More details on the generation process are given later in the methods section. Note that the body sizes of the chosen bodies vary greatly as, e.g., Bennu has a diameter of merely 525 m while Eros has one of almost 17 km. For convenience and consistency we therefore introduce and use non-dimensional units for all the obtained models. As a result, the integration volume $V$ for all cases is reduced to be the cube $[-1,1]^3$ which is ensured to contain all of the asteroid mascons.

\subsection*{Learned models}
We apply the training pipeline depicted in Figure (1) to obtain a neural density field for each of the nine mascon models generated including the heterogeneous models for Bennu, Itokawa and Planetesimal. The quality of the final result is assessed quantitatively by comparing the ground-truth acceleration from the mascon model to the accelerations caused by the neural density field at 10,000 random validation points located at specific altitudes.
For each body we fix a low, medium and high altitude corresponding respectively to 0.04, 0.08 and 0.2 units of length. An example of the resulting validation points in the case of Eros is shown in Figure \ref{fig:validation_points} for the medium altitude case.
Additional details on the training, validation and sampling are given in the methods section.
In Table \ref{table:main_results} we show the results in terms of mean absolute error and mean relative error. 
For all cases considered, our geodesyNet is able to learn the mass density in the volume $V_B$ such that it reproduces the ground-truth gravity field with a relative error between 0.12\% and 0.89\% on all bodies at all tested altitudes. Unfortunately, a direct quantitative comparison with prior literature is not viable, as a common validation practice has not been established and, further, the assumptions on the information available on the bodies made in other works differ (here no knowledge on the shape model is assumed). The closest candidate for a first, preliminary, comparison is perhaps, in our case, the work by \emph{Wittick and Russell} \cite{wittick2019mixed} who provide a detailed analysis on the performance of a state-of-the art method representing the gravity field of Eros, albeit informed by a shape model. They make use of a hybrid model joining mascon and spherical harmonics modelling. A first comparison to their work is provided in the supplementary materials, Section S1.

\begin{figure}[ht]
    \centering
    \includegraphics[width=1.0\columnwidth]{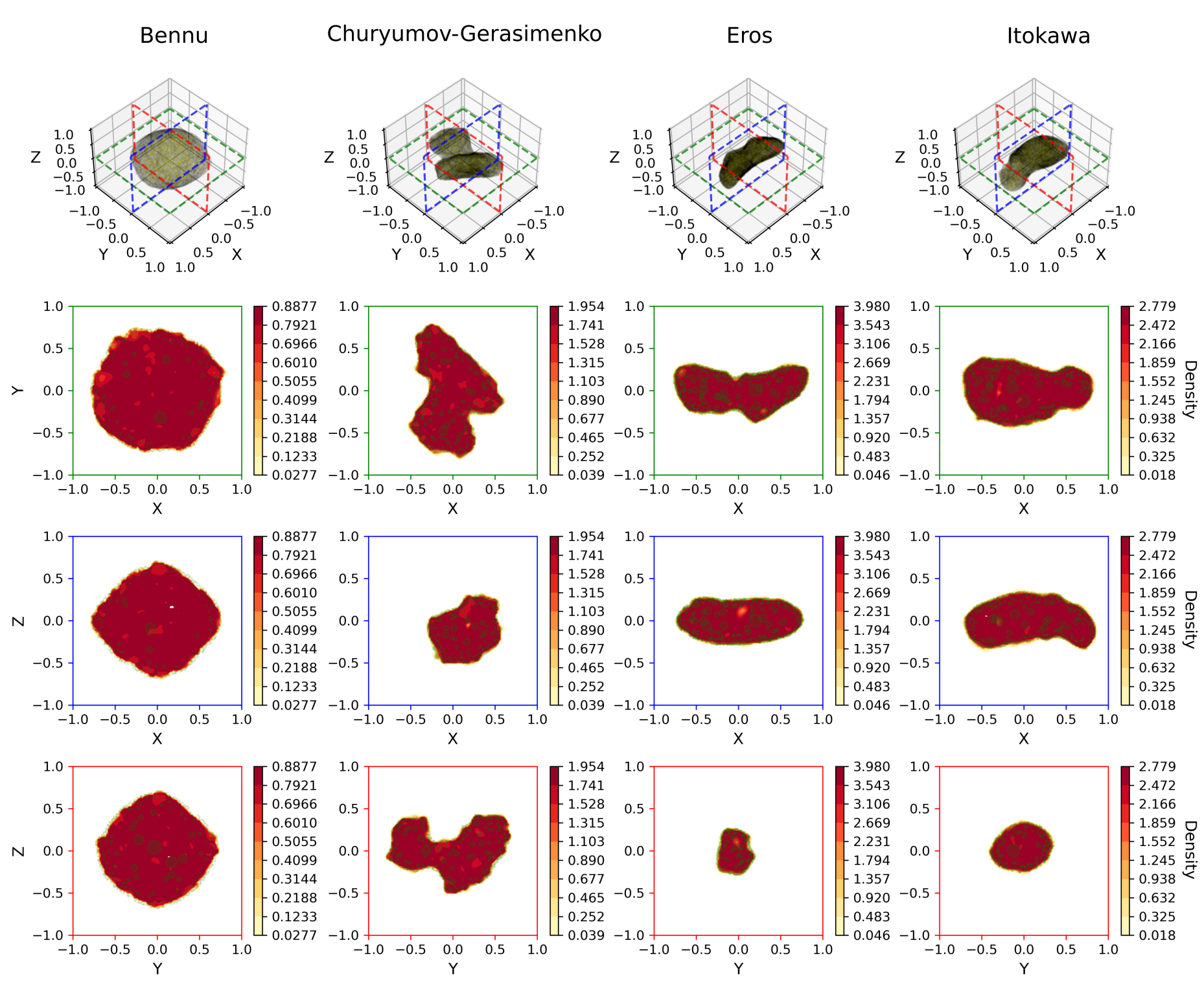}
    \caption{Results on the four (homogeneous) solar system bodies. Ground-truth models are in the first row as well as the selected slice location for the slices depicted below. For each slice a heatmap of the neural density field is overlaid with the mascons (green). Only mascons within a small distance to the selected slice are shown, hence the apparent mascon sparsity. The neural density fields are able to reconstruct correctly the body shape as well as their internal homogeneous density. \label{fig:homogeneous}}
\end{figure}

\begin{figure}[ht]
    \centering
    \includegraphics[width=1.0\columnwidth]{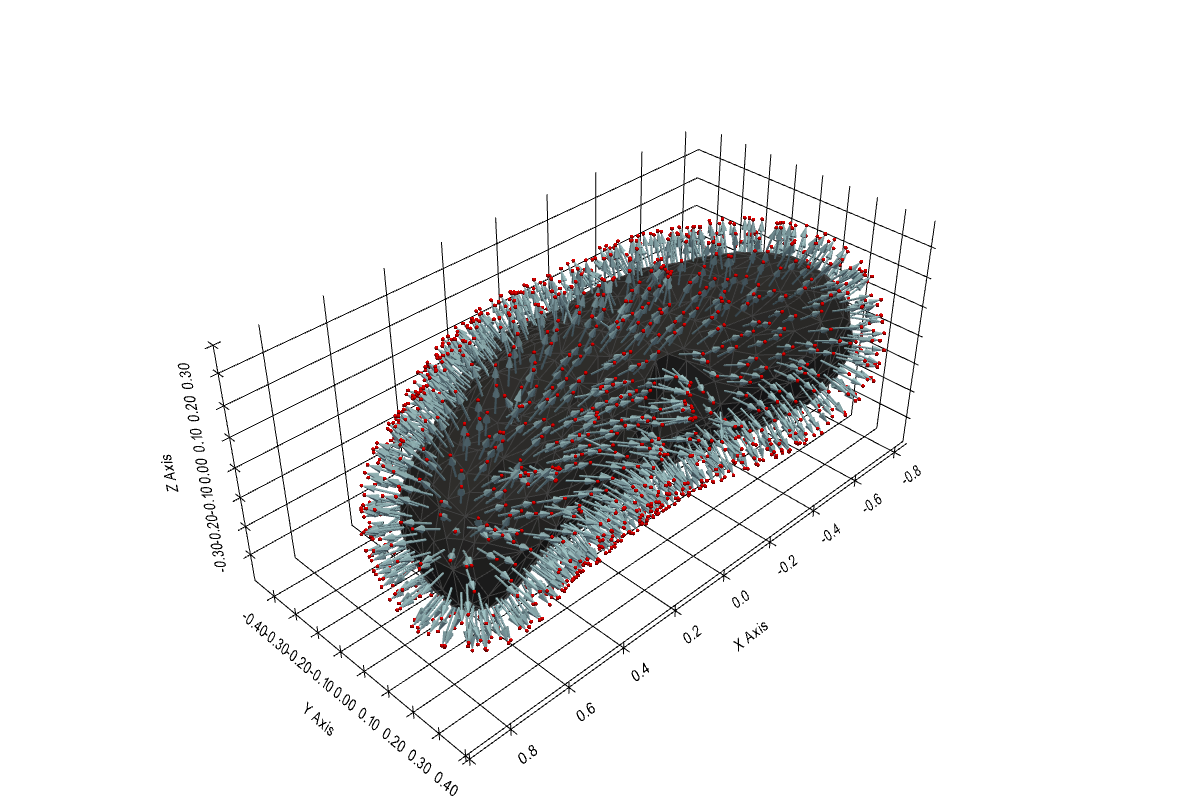}
    \caption{Visualization of the validation point sampling on the example of Eros. Points (red) are sampled at specific altitude - 0.1 in  non-dimensional units, equivalent to 2042.5m - using the normal vectors (blue) of the body mesh triangles (grey). Points with mismatching altitude due to the non-convex nature of the mesh are automatically discarded. Additional details are given in the Methods Section. \label{fig:validation_points}}
\end{figure}

\begin{figure}[ht]
    \centering
    \includegraphics[width=0.9\columnwidth]{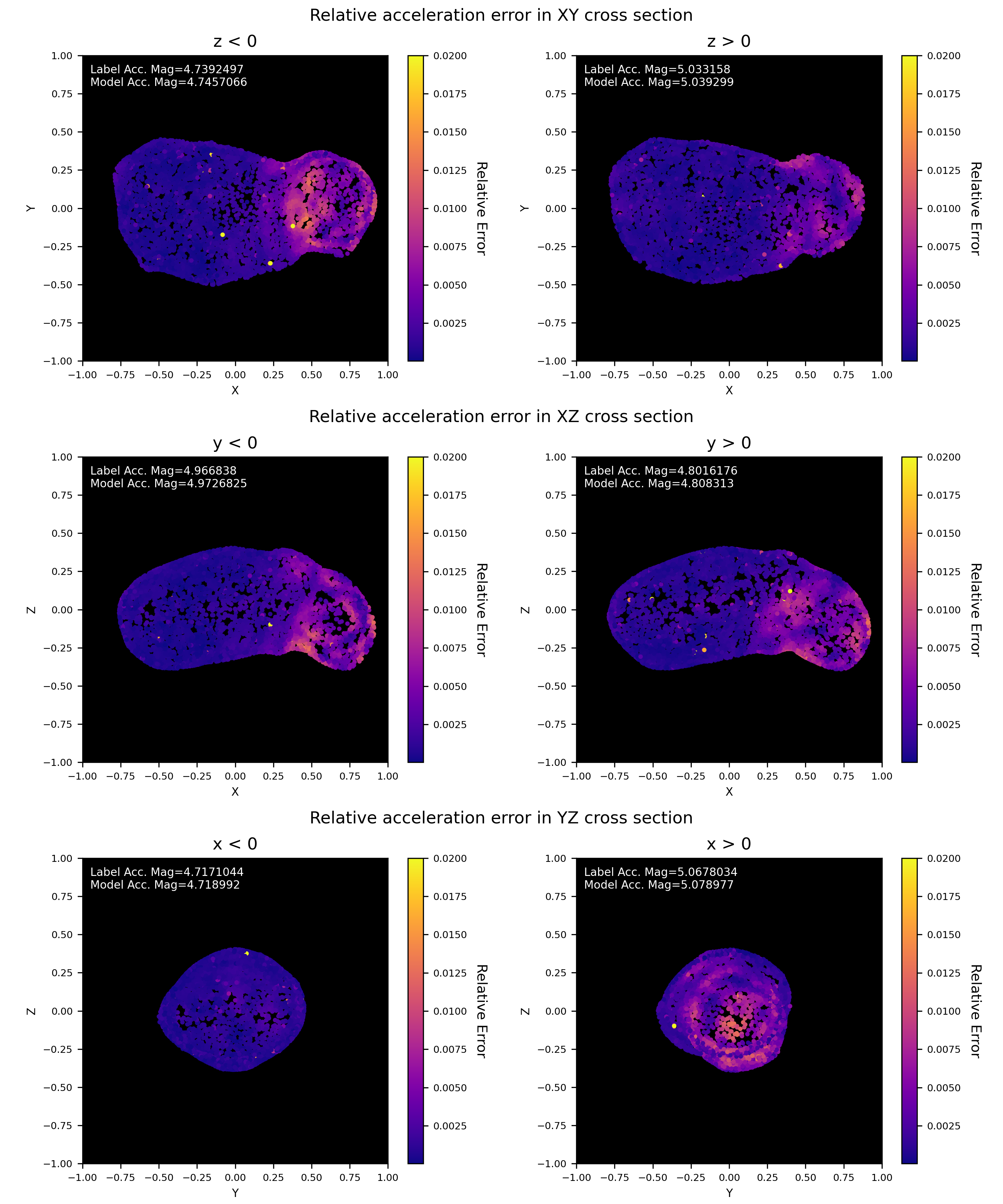}
    \caption{Visualization of relative errors of the gravitational acceleration predicted by the trained geodesyNet close to the body's surface. Points are located 17.5m above the asteroid surface (0.05 in non-dimensional units). The spread is quite uniform, except for points closer to the heavier asteroid head that are correlated to higher errors, as expected. \label{fig:acc_err}}
\end{figure}

A visual indication of the achieved accuracy for all the homogeneous solar system bodies is given in Figure \ref{fig:homogeneous}, where the neural density field is plotted against the mascon ground truth. Note that even small-scale surface features such as larger rocks and craters are reconstructed to some extent. 
For completeness, we also report the relative error close to the surface, in the case of the heterogeneous Itokawa model, in Figure \ref{fig:acc_err}. The error distribution across the asteroid surface appears overall to be quite uniform along the asteroid body, revealing how the neural density field is able to balance errors caused by the presence of complex surface features. It is worth to note how most of the relative error is, as expected in the case of the heterogeneous Itokawa, concentrated around the heavier asteroid head where the close proximity to the larger source of the gravitational field is penalizing.

\begin{table}[tb]\scriptsize
\caption{Learned models -- mean absolute and relative acceleration errors at three altitude (low, medium, high). Altitudes for validation are chosen depending on the body size as a fraction of its diameter. HMG abbreviates homogeneous, and HTG heterogeneous, respectively. \label{table:main_results}}
\begin{center}
 {\tabulinesep=1.5mm
 \setlength\tabcolsep{2pt}
\begin{tabu} { | X[0.4cm] |X[1.125m]||  X[0.9cm]|  X[0.9cm]|  X[0.9cm]|| X[1cm] | X[1cm] | X[1cm]|| X[0.675cm] | X[0.675cm] | X[0.675cm] |}
 \hline
   & & \multicolumn{3}{c||}{Sampling Altitudes} & \multicolumn{3}{c||}{Absolute Errors} & \multicolumn{3}{c|}{Relative Errors}\\
 \hline
    & Body & $h_{low} [m]$ & $h_{med} [m]$ & $h_{hi} [m]$ &$\epsilon_{low} [m/s^2]$ & $\epsilon_{med} [m/s^2]$ & $\epsilon_{hi} [m/s^2]$ & 
           $\epsilon_{low} [\%]$ & $\epsilon_{med} [\%]$ & $\epsilon_{hi} [\%]$ \\
 \hline
\multirow{6}{*}{\rotatebox[origin=c]{90}{\scriptsize HMG}} & Bennu 			        &   14.1 &   28.2 &   70.4 &   8.92E-08 &   4.81E-08 &   1.88E-08 & 0.371 & 0.223 & 0.116 \\ \cline{2-11}
& Churyumov-Gerasimenko 	&   125 &   250 &   625 &   3.33E-07 &   1.70E-07 &   5.56E-08 & 0.556 & 0.319 & 0.146 \\ \cline{2-11}
& Eros 			        &   817 &   1630 &   4080 &   6.82E-06 &   3.22E-06 &   9.96E-07 & 0.467 & 0.259 & 0.123 \\ \cline{2-11}
& Itokawa 			    &   14 &   28 &   70.1 &   8.99E-08 &   4.28E-08 &   1.57E-08 & 0.407 & 0.225 & 0.124 \\ \cline{2-11}
& Planetesimal 			        &   125 &   250 &   625 &   1.24E-07 &   7.87E-08 &   3.52E-08 & 0.423 & 0.293 & 0.168 \\ \cline{2-11}
& Torus 			        &   125 &   250 &   625 &   4.84E-07 &   2.58E-07 &   9.19E-08 & 0.891 & 0.577 & 0.318 \\ \hline \hline
\multirow{4}{*}{\rotatebox[origin=c]{90}{\scriptsize HTG}} & Bennu  &   14.1 &   28.2 &   70.4 &   1.17E-07 &   6.22E-08 &   2.71E-08 & 0.473 & 0.279 & 0.163 \\ \cline{2-11}
& Itokawa &   14 &   28 &   70.1 &   8.94E-08 &   4.17E-08 &   1.46E-08 & 0.407 & 0.219 & 0.113 \\ \cline{2-11}
& Planetesimal &   125 &   250 &   625 &   1.19E-07 &   7.90E-08 &   4.08E-08 & 0.349 & 0.255 & 0.172 \\ \hline
\end{tabu}}
\end{center}
\end{table}

\subsection*{Adding information on the body shapes}
\begin{figure}[ht]
    \centering
    \includegraphics[width=1.0\columnwidth]{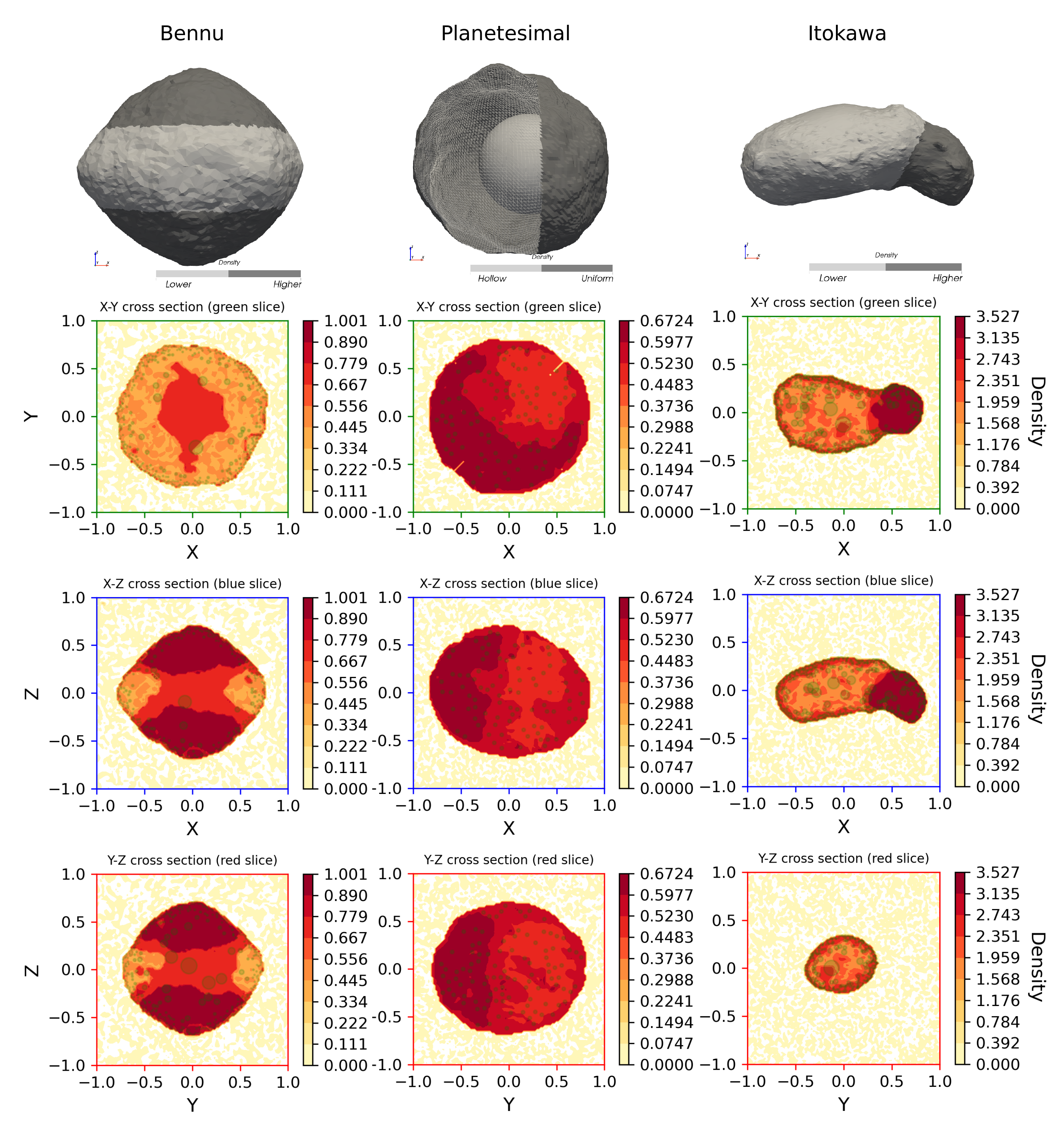}
    \caption{Differential training -- Results on the three heterogeneous bodies. Ground-truth models are in the first row indicating the lower and higher density regions. For each slice a heatmap of the reconstructed neural density field is overlaid with the mascons (green). Only mascons within a small distance to the selected slice are shown, hence the apparent mascon sparsity.\label{fig:nu}}
\end{figure}

If additional information on the body shape $\partial V_B$ is available, this can be seamlessly integrated into the geodesyNet training to improve the results. A geodesyNet can in fact be trained to learn, instead of the body density, the density variation from a homogeneously distributed density inside $\partial V_B$. We refer to this variant of the training pipeline as differential training (see more details in the methods section). We apply the differential training pipeline only to the three heterogenous mascon models, as in the homogeneous cases the neural density field learned by the differential approach is trivial as it would vanish entirely and so would the obtained relative errors.
Figure \ref{fig:nu} visualizes the heterogeneous ground truths and the network's predicted distributions.
In Table \ref{table:diff_results} the relative and absolute errors on the predicted accelerations are given. In comparison to the previously trained geodesyNets (see Table \ref{table:main_results}), differential training achieves better results in terms of acceleration error reducing it consistently up to a factor of four. Relative errors all lie between 0.1\% and 0.25\% on all samples. \\
In the heterogeneous Bennu case, the neural density field clearly shows the presence of a lower density region at the equator (see Figure \ref{fig:nu} left column) and it also outputs density values compatible with the ratio $f=2$ used to create the ground-truth density difference between polar and equatorial regions.
Note that the asteroid core is left at a slightly higher density than the outer layers at the equator -- a solution selected by the network bias among the ones compatible with the observations.
For the heterogeneous version of Itokawa, the trained geodesyNet is able to precisely reconstruct the higher density region at the asteroid's \lq\lq head\rq\rq\, closely matching the ground truth (see Figure \ref{fig:nu} right); also from a quantitative point of view as the predicted density values are compatible with the correct ratio $f=1.6$. A small inaccuracy is notable in the predicted slightly higher density of a thin layer close to the surface, likely a numerical artefact caused by the mismatch between the shape of the asteroid and the mascon approximation near the surface. Mascon models suffer from inaccuracies next to the body surface where their point mass nature becomes evident.
For the case of Planetesimal, while differential training improves the description of the surrounding gravity field and reduces the relative error on the predicted relative acceleration from 0.34\% to 0.096\% in the near field -- see Tables (\ref{table:diff_results}) and  (\ref{table:main_results}) -- we note that the obtained neural density field fails to fully reproduce the cavity inside the asteroid, settling instead for a solution where the ratio between the density inside the cavity and the one outside is not zero. We must note that the gravity inversion problem is, for this particular shape, particularly challenging as the heterogeneity chosen for this rather symmetrical Planetesimal nears a configuration where the Shell theorem would apply, exacerbating the ill-posed nature of the gravity inversion problem \cite{cicci1992improving, tricarico2013global, park2010estimating, russell2012global}. 

\begin{table}[tb]\scriptsize
\caption{Differential training -- mean absolute and relative acceleration errors for  heterogeneous bodies at three altitude (low, medium, high). latitudes for validation are chosen depending on the body size as a fraction of its diameter.}
\begin{center}
 {\tabulinesep=1.5mm
 \setlength\tabcolsep{2pt}
\begin{tabu} { |X[1.125m]||  X[0.9cm]|  X[0.9cm]|  X[0.9cm]|| X[1cm] | X[1cm] | X[1cm]|| X[0.675cm] | X[0.675cm] | X[0.675cm] |}
 \hline
  & \multicolumn{3}{c||}{Sampling Altitudes} & \multicolumn{3}{c||}{Absolute Errors} & \multicolumn{3}{c|}{Relative Errors}\\
 \hline
  Heterogeneous body & $h_{low} [m]$ & $h_{med} [m]$ & $h_{hi} [m]$ &$\epsilon_{low} [m/s^2]$ & $\epsilon_{med} [m/s^2]$ & $\epsilon_{hi} [m/s^2]$ & 
          $\epsilon_{low} [\%]$ & $\epsilon_{med} [\%]$ & $\epsilon_{hi} [\%]$ \\
 \hline
Bennu 			     &   14.1 &   28.2 &   70.4 &   5.97E-08 &   2.76E-08 &   1.23E-08 & 0.247 & 0.126 & 0.075 \\ \hline
Itokawa 			 &   14 &   28 &   70.1 &   3.71E-08 &   1.97E-08 &   1.20E-08 & 0.166 & 0.104 & 0.094 \\ \hline
Planetesimal 			 &   125 &   250 &   625 &   3.25E-08 &   1.62E-08 &   1.16E-08 & 0.096 & 0.053 & 0.049 \\ \hline
\end{tabu}}
\label{table:diff_results}
\end{center}
\end{table}

\section*{Discussion}
\subsection*{Gravity representation}
Prior to our work, three main representations of a gravity field have been widely studied and used in the context of geodesy: spherical harmonics, mascon models and polyhedral gravity representations \cite{paul1974gravity}. A qualitative comparison with these approaches is displayed in Table 3.
\subsubsection*{Spherical harmonics}
The spherical (or the spheroidal) harmonics approach allows the representation of a generic gravity field via the coefficients of its Fourier series expansion in spherical (or similar) coordinates. Its use is particularly suited for bodies, such as large planets and moons, that have a strong axial symmetry and regularity. 
For example, thanks to accurate data collected during missions such as the Steady‐state Ocean Circulation Explorer (GOCE) \cite{drinkwater2003vii} or the Gravity Recovery and Climate Experiment (GRACE) \cite{tapley2004grace} spherical harmonics expansions could be computed to describe the Earth's gravity field up to wavelengths of the order of $\approx 160$ km \cite{visser1999gravity}, including terms of degree up to 250. However, applying the same technique to achieve comparable precision for irregular shaped bodies is possible but troublesome. 
Inside the Brillouin sphere \cite{takahashi2014small}, the convergence of spherical harmonics expansions is known to be erratic \cite{hirt2017convergence} and requires special attention \cite{sebera2016spheroidal}.
Outside the Brillouin sphere the convergence becomes slower as soon as the body shape has irregularities departing significantly from a simple reference triaxial ellipsoid. 
In the case of small solar system bodies such as comets and asteroids, the use of such an expansion outside the Brilloiun sphere, when possible, requires an increasingly high number of coefficients.
For example, \emph{Seber et al.} \cite{sebera2016spheroidal} developed a spheroidal harmonics model for the relatively regularly shaped asteroid Bennu using terms of degree up to 360 and reported relative errors, on the obtained potential, of the order of $\approx 1\%$ in the exterior field (e.g. at 5-20m from the Brillouin sphere). It has been proposed to make use of a different expansion for the interior gravity \cite{takahashi2013surface} at the cost of introducing an artificial boundary at the Brillouin sphere surface. 
A geodesyNet model makes no distinction between inner and exterior field as it equally learns to match gravitational observations inside and outside of the Brillouin sphere. Its neural architecture, based on the neural radiance fields, is not impacted as much by the complexity of the represented bodies as shown also by the work of \emph{Mildenhall et al.} \cite{mildenhall2020nerf} that reports success in encoding many different three-dimensional scenes using a fixed amount of parameters.

\subsubsection*{Mascon Models}

A second method classically used to represent the gravitational field is that of filling the volume occupied by the body with point masses or \lq\lq mascons\rq\rq. 
Each mascon is assigned a mass so that the total body mass is reconstructed. All our ground-truth gravity fields were generated this way.
The approach has the great advantage of its simplicity and it is straightforward to implement, but it also has several deficiencies when employed to represent observed gravity fields.
The number of mascons needed to achieve accuracies comparable to that of spherical harmonics in the description of the exterior field is very large \cite{werner1996exterior}. 
In the interior field, and specially close to the surface, the accuracy is troublesome as it becomes unclear whether a point is underneath or just over the body surface. These drawbacks can be reduced using hybridized techniques \cite{wittick2019mixed} and introducing knowledge on the shape and composition of the asteroid. 
A geodesyNet, as shown in the result section, is able to achieve comparable performances to that of mascon hybridized techniques while using no prior information on the body shape and returning a continuously differentiable representation of the internal body density.

\subsubsection*{Polyhedral Gravity}

A last approach, polyhedral gravity, has been widely used in the context of irregular bodies' gravity. Therein, the divergence theorem, or Gauss's theorem, is used to transform the triple volume integral needed to compute the gravitational acceleration or potential into the integral along the asteroid surface of its flux\cite{paul1974gravity}, \cite{werner1996exterior}. The surface is then approximated by a polyhedron and an analytical formula is derived that enables computing the gravity produced by any polyhedral body having a homogeneous density. This technique is only valid for homogeneous bodies and requires availability of a high-fidelity model of the body shape to derive a polyhedral model from. Often, this is computed on-ground thanks to reconstruction techniques based on the images available from on-board cameras. This method can also have a high computational complexity as a large amount of evaluations of functions like arctangents and logarithms is needed \cite{d2014analytical}. A geodesyNet, on the other hand, does not need a homogeneous density body to be able to learn a representation of the gravity field, but it is able to use, if available, the shape information to improve the quality of its predictions and to build plausible models for the internal structure of heterogeneous asteroids. In Table \ref{table:comp} we have summarized the different properties of fundamental representations of a gravity field, highlighting the broad applicability and various use cases of the proposed geodesyNets.

\begin{table}[t]
\caption{Comparative table of different fundamental approaches able to represent and learn gravity fields of three-dimensional bodies.\label{table:comp}}
\centering
\begin{tabular}{|l|c|c|c|c|}\hline
 & \multicolumn{4}{c|}{Approach} \\ \hline
Characteristic                      & Mascons   & Harmonics & Polyhedral    & geodesyNets   \\\hline
Differentiable                      & no        & yes       & yes           & yes           \\\hline
Viable inside Brillouin sphere      & yes       & no        & yes           & yes           \\\hline
Allows for heterogeneous density    & yes       & yes       & no            & yes           \\\hline
Requires shape model                & yes       & no        & yes           & no            \\\hline  
Can utilize shape model             & yes       & no        & yes           & yes           \\\hline  
Accurate in the near field          & no        & yes       & yes           & yes           \\\hline
\end{tabular}
\end{table}

\subsection*{Gravity inversion}
From a methodological viewpoint, there are several distinguishing factors and relevant parallels of geodesyNets in comparison to previous approaches. One noteworthy aspect of geodesyNets is the ability to serve several purposes at once: after training, the same geodesyNet can be used to represent both the gravitational field around a body as well its shape and density. In detail, the geodesyNet learns a representation for the gravitational field outside of the body, for the body density inside -- i.e. the gravity inversion problem -- and for the body shape $\partial V_B$ itself. This stands in contrast to past approaches, for example, on gravity inversion - either employing a mascon perspective \cite{russell2012global} or working in the mass density space directly \cite{chambat2005empirical, tricarico2013global, berkel2010mathematical} - which rely on the existence of a shape model for the body. \par
There are also notable similarities with other approaches. Previous approaches assume a kernel and create a highly parametric model of the body density inside a known volume and then fit it to reproduce gravitational observations \cite{tricarico2013global}. This is to some degree similar to our approach, where the kernel functions are, analogously, defined by the neurons' non-linearities and the network weights are the model parameters. Stochastic gradient descent then allows fitting the model parameters to the observations. From this viewpoint, the main difference between past gravity inversion methods and the use of a geodesyNet stems from the mathematical properties of the parametric representation employed. \par
Unlike previous approaches we use a feedforward artificial neural network and thereby rely on its universal function approximator property (see \emph{Calin} \cite{calin2020universal}). This property makes it particularly well suited to describe sharp density discontinuities, such as those encountered when crossing $\partial V_B$ or across possible interfaces between density layers. A more traditional polynomial or spline \cite{berkel2010mathematical} parametrization, while in principle able to capture these effects, would require far too many coefficients to capture this feature with similar precision. In consequence, such density representations are typically limited to a given, known volume within the body \cite{tricarico2013global} and are unable to represent the whole geodetic properties of a given body. Overall, geodesyNets replicate the results from previous approaches while providing a more holistic solution which requires fewer assumptions.

\subsection*{On-board utilization}
The accurate characterization of the spacecraft orbital environment is a crucial requirement of missions to comets and asteroids \cite{park2010estimating}. The approach we introduce here offers a potential simplification to this critical mission phase if the training of a geodesyNet can be performed on-board and in real-time while the spacecraft performs its orbits around the target body. This might save mission resources by eliminating the need to collect visual information for, e.g., a three-dimensional reconstruction of the shape. Further, if performed on-board the shape information can be collected online and less data may have to be transmitted.
We argue that such a possibility, while currently not fully developed, is likely to become available in the near future. The use of dedicated on-board hardware enabling advanced artificial intelligence approaches for space missions has been recently reviewed by \emph{Furano et al.} \cite{furano2020towards} who discusses radiation hardened GPUs as well as FPGAs or hardware accelerators such as Myriad 2 \cite{moloney2014myriad}. 
In our case, as more batches of data would become available to the on-board computer, these could be seamlessly exploited by continuously adjusting the network parameters with some update rule to gradually improve the neural density field stored in the on-board geodesyNet.
The network would be able to continuously learn during various mission phases -- also accounting for unforeseen deviations and anomalies in the incoming data -- and eventually be sent back to the ground as a compressed, differentiable, representation of the body shape and its first plausible internal structure. 
The memory requirement during training is perhaps the main limiting factor for the precision achievable in a possible on-board utilization, being mainly driven by the number of points $N$ used for the numerical quadrature used to evaluate the volume integral in Eq.(\ref{eq:quadrature}). Adjustments to enable a smaller batch size or a more memory-efficient numerical integration scheme may alleviate these concerns, however. On the other hand, the number of model parameters -- see supplementary material Table S4 -- is relatively modest and at most a few hundreds of kilobytes are needed to store a trained geodesyNet. 
Note that reasonable accuracy for a rough estimate of the density distribution and induced acceleration is obtained even when using a small number of parameters. 

\section*{Methods}

The training process shown in Figure 1 is merely an outline and a number of important details were found to be important in training a geodesyNet. Hence, we describe the training setup in more detail and focus in even more depth on the choice of loss function, the numerical integration method and the differential training. 

\subsection*{Training Setup}
At its core the chosen neural network architecture is reminiscent of a SIREN network as proposed by Sitzmann \etal{} \cite{sitzmann2020implicit}. It is a fully connected network with nine hidden layers of 100 neurons with sinusoidal activation functions in between layers. The final activation layer computes an absolute value (densities should be positive) or - in case of the differential training - a hyperbolic tangent (densities variations can be signed). A detailed analysis of the impact of the architecture details ( SIREN's $\omega$, layers, neurons) is given in the Supplementary (Tables S1, S2, S5).\\
The network is trained using the ADAM optimizer with an initial learning rate of $1^{-4}$. The loss function has been specifically designed for geodesyNets and is presented in the next section. During training, a learning rate decay is used when the training loss plateaus for 200 iterations with a reduction with a factor of 0.8 up to a minimum learning rate of $1^{-6}$. Anecdotally, we observed several reductions in most training runs. The batch size during training is comparatively high at 1000 to average out the impact of noise in the form of numerical errors during the numerical integration. As a larger number of parameter studies is part of this work requiring hundreds of GPU hours, only one random seed was investigated for each run. Training consists of up to 10,000 iterations with early stopping on the training loss after a warmup of 3,000 iterations, if 2,000 iterations without a new optimum are computed. The model with optimal training loss is then used for validation. \par
The synthetic observations which serve as the network's training data require a choice of sampling points at which the ground-truth acceleration is computed. The choice is critical to ensure numerical stability and plausibility (in practice only sampling outside the body is plausible for spacecraft) while optimally providing information to the network. Hence, sampled points were sampled uniformly inside a unit sphere around the center of the body but always outside the body (the coordinate frame is chosen so that the body fits precisely inside $[-0.8,0.8]^3$). The location of the sampled points (inside / outside the body) is determined with a custom PyTorch implementation of the ray-triangle intersection algorithm  by M\"oller \& Trumbore \cite{moller1997fast}. A low-poly version of the body meshes is used for the location determination. In a practical application the sampled points would not be random, but be determined by, e.g., a spacecraft's trajectory or observed particles' locations. For computational efficiency, we resample new observations only every ten training iterations. Numerical integration and a mascon model then provide the ground-truth label as described in the previous sections. The network's prediction is analogously computed using the numerical integration over the neural density field. The method for the numerical integration is explored in more detail in the following sections. \par
To compute robust results on previously unobserved data we implemented a high-accuracy validation procedure. The validation consists of computing all reported error values on, respectively, 10,000 randomly sampled points at three specific altitudes above the body (0.04, 0.08, 0.2 length units). The sampling is displayed in Figure 5. The validation utilized a larger number (500,000) of samples in the numerical integration and high-fidelity - not low-poly - meshes for the location ( inside / outside) determination.  

\subsection*{Loss Functions}
As loss function for a geodesyNet we investigated several standard machine learning losses such as the Mean Squared Error or the Mean Absolute Error. While such choices would already result in successful training, especially for regular body shapes, they can be considerably improved by assuming the network predictions $\hat y_i$ as biased by a factor $\kappa$. Then, given a batch of $N$ ground-truth quantities $y_i, i=1..n$ and predictions  $\hat y_i, i=1..n$, over a batch, the optimal value of $\kappa$ can be found to be:
$$
\kappa = \frac{\sum_{i=1}^n \hat y_i y_i}{\sum_{i=1}^n y_i^2}
$$
This normalizing factor is further referred to as mass normalization factor and scales the neural model predictions so that the value $\kappa\hat y_i$ is used in the loss function. In the case of the Mean Absolute Error, this produces the new loss function:
$$
\mathcal L_{\kappa MAE} = \frac  1n \sum_{i=1}^n{\left|y_i - \kappa \hat y_i\right|} = \frac  1n \sum_{i=1}^n{\left|y_i - \frac{\sum_{i=1}^n \hat y_i y_i}{\sum_{i=1}^n y_i^2} \hat y_i\right|}
$$
The mass normalization factor has the effect of ensuring that the overall asteroid mass, i.e. the integral over the volume $V$ of the density predicted by the geodesyNet, is set as to minimize the resulting mean squared error. This adds one free parameter to the model, but a parameter that is analytically found at each step from the batch data. 
This way, the error backpropagation can focus on learning the deviation from a homogeneously filled volume $V$, and not, concurrently, find the absolute value of the body mass. In this work, we experimented with a number of different losses (supplementary Table S5) with the $\kappa MAE$ being the final choice for the best performing models.

\subsection*{Numerical Integration}
To compute the acceleration $\mathbf a(\mathbf r)$ induced by some density distribution $\rho(\mathbf x)$ (e.g. from the last layer of the geodesyNet) we use the formula:
\begin{equation}
\label{eq:quadrature}
   \mathbf a(\mathbf r) = G \int_{\mathbf x \in V} \frac{\rho(\mathbf x)}{|\mathbf r - \mathbf x|^3}\left(\mathbf r - \mathbf x\right) dV.
\end{equation}
where $G$ is the Cavendish constant. Due to the lack of analytical solutions, these integrals have to be computed numerically in such a way as to allow for the error to be backpropagated through the integral. Given that the integral for the acceleration is a triple integral, getting a sufficiently accurate approximation is the computationally most expensive factor during the network training. For the computation of the integral approximation the network has to be evaluated at a large number of points (in our case $3$ to $5 \cdot 10^5$ points). At the time of the experiments no public code was available to compute more than one-dimensional integrals while maintaining automatic differentiation capabilities on a GPU. Hence, we implemented two methods for this problem ourselves.  We tested two methods for solving the necessary integral numerically, a Monte Carlo Integrator \cite{caflisch1998monte} and a classical composite trapezoidal rule. Comparatively simple methods were chosen as a full vectorization and implementation in \textit{PyTorch} was necessary for computational efficiency and to enable backpropagation through the integration. We showcase the convergence properties in the supplementary material in Figure S1.
In practice, each training iteration of the network requires the evaluation of the integral for all sampled points $r$. However, the evaluation of the network $\rho(\mathbf x)$ has to be performed only once per iteration. Thus, an efficient implementation of the numerical integration algorithm is paramount. With the chosen implementation, our integration methods demonstrated absolute errors with magnitude $\approx 1e-4$. Even though neural networks are somewhat robust to noise (such as numerical errors), this is likely a limiting factor for the network's obtainable accuracy and we suspect that improving the numerical accuracy of the integral computation may enable even better results.

\subsection*{Differential Training}
To improve the resulting neural density fields for heterogeneous bodies we developed a separate training procedure relying on the knowledge on the body shape information. Without such information the neural density field, while still able to reproduce the observed gravity measurements within a good approximation (see Table \ref{table:main_results}), tends to create an homogeneous body contained in $\partial V_B$. This innate bias of networks towards low frequencies discussed by \emph{Rahaman et al.} \cite{rahaman2019spectral} is unavoidable as the loss function is not able to distinguish between the two possible and equally valid solutions -- a consequence of the fact that the gravity inversion problem is ill-posed. In those cases where the body shape is known, though, the network can, and should, be informed on what areas of the volume $V$ do contain a vanishing density and which ones don't. A simple, albeit elegant, solution to indirectly introduce this information in the loss is to train the geodesyNet to learn the difference between the measured acceleration $y_{nu}$ and that created assuming a perfectly homogeneous body $y_u$ -- which we can compute since the body shape is assumed as known. 

In our experiments on differential training, consistently with the rest of our setup, we used mascon models to compute both $y_u$ and $y_{nu}$, but it is worth mentioning here that in different setups, the computation of $y_u$ could also be done using a polyhedral gravity approach which is an exact solution for a polyhedral shaped body. The geodesyNet is then tasked with predicting the local density difference between the bodies that causes the difference in acceleration. Thus, for each sampling point $r_i$, based on ground-truth acceleration $y_u$ and $y_{nu}$ for the homogeneous and heterogeneous, respectively, we optimize the network parameters $\theta$ to minimize

\begin{equation}
    \mathcal L_{\kappa MAE} = \frac  1n \sum_{i=1}^n \left| y_{nu} - y_u - \kappa \tilde{y} \right|.
\end{equation}
Note that this implies the availability of a homogeneous density model of the target body, and hence knowledge of its shape. In practice, this may be obtainable even on a spacecraft as shown by Bandyonadhyay \etal{} \cite{bandyonadhyay2019silhouette}. Then, the expected homogeneous acceleration $y_u$, would be precomputed from a polyhedral gravity model, for example, while  $y_{nu}$ is observed during flight.

\subsection*{Generating the mascon models ground truths}
In the case of the asteroids 433 Eros, 101955 Bennu and 25143 Itokawa and the comet 67P Churyumov–Gerasimenko, we obtain the high-fidelity polyhedral model shapes reconstructed from the various instruments on-board spacecraft that visited the asteroid: for Eros and Itokawa we use the models produced by \emph{Robert Gaskell} \cite{erospoly} \cite{itokawapoly}, for Bennu we use the model made available by the Osirix-REX team \cite{bennupoly} and for Churyumov-Gerasimenko we use the model made available by the European Space Agency \cite{67ppoly}. 
Since the polyhedral models of the asteroid surface are not suitable to represent the gravitational field of heterogeneous bodies, we transform the surface meshes into mascon models first creating a constrained Delaunay tetrahedralization \cite{si2015tetgen} and then placing a mass $m_j$ at the centroid of each resulting tetrahedron.

To further add variety to our dataset, we generate two additional mascon models not representative of any real body in the solar system:  Planetesimal and Torus. 
The model for Planetesimal is obtained from N-Body simulations made at the Max Planck Institute for Astronomy using the SMC paradigm \cite{chrono_smc} and aimed at reproducing planetesimal formation. 
A final, statically stable planetesimal, is extracted from the simulation and used directly. To obtain Torus, instead, we create, similarly to what was done for the Solar System bodies, a mascon model from a starting polyhedral mesh representing a toroidal object. Both Planetesimal and Torus are then arbitrarily assumed to have the same mass and diameter as that of the comet Churyumov–Gerasimenko.
For all bodies, we consider the homogeneous case by setting the values of the masses $m_j$ to a value proportional to the corresponding tetrahedron volume and taking care that the sum of all masses reconstructs the actual body mass. For the planetesimal case we simply set all values as equal since the body is itself a stable aggregate of spherical masses.

For Bennu, Itokawa and Planetesimal we also generate models having a heterogeneous mass distribution. In the case of Bennu, flight data from the Osirix-REX mission indicated a possible area with a lower density at the equator \cite{scheeres2020heterogeneous}, hence we generate the heterogeneous version of the model with a fictitious higher density area in the polar regions. We multiply the values of the mascon masses taken from the homogeneous model and belonging to the polar regions, by an arbitrary factor $f=2.0$ and we renormalize the overall asteroid mass. Similarly, in the case of Itokawa, flight data revealed a possible higher density in the rubble pile head \cite{fujiwara2006rubble}. We thus create a heterogeneous version of the mascon model by multiplying all mascons  masses taken from the homogeneous model and belonging to the asteroid head by a factor $f=1.6$, in agreement with the data published by Lowry et al. \cite{lowry2014internal}, and normalizing again the overall asteroid mass. In the case of the Planetesimal, we generate a hollow structure by setting all mascon masses belonging to an internal spherical region to zero, and renormalizing the remaining mascon masses.

To standardize all our experiments and be able to develop a unique numerical pipeline for all the very different bodies here studied, we introduced non-dimensional units for the length, the mass and time. In particular, we set the integration volume $V$ in Eq.(\ref{eq:quadrature}) to be the hypercube $[-1,1]^3$ and rescale the body mascon model so that the maximum absolute value of its coordinates is $\delta_{max} = 0.8$. Hence we derive the unit length $L$ for the body. We then set the unit of mass to the body mass $M$ and derive the value of the units of time by setting the Cavendish constant $G$ to one. The resulting units as well as other parameters of the various models used for each body are shown in the supplementary materials Table S3.

\bibliography{scibib}

\bibliographystyle{ScienceAdvances}

\section*{Acknowledgments}
The authors are grateful to Dr. Francesco Biscani from the Max Planck Institute of Astronomy (Heidelberg) for making available stable configurations of plausible planetoids to be used as mascon ground truths in the paper and to Dr. Dawa Derksen for the interesting discussions and exchanges on neural scene representations.

\section*{Authors Contribution}
Dario Izzo formulated and led the project. Dario Izzo and Pablo \gomez refined the methodology and developed the code base. Pablo \gomez automated and performed the numerical experiments. Dario Izzo and Pablo \gomez wrote and revised the paper.

\section*{Competing interests} 
The authors declare that they have no competing interests. 

\section*{Data and materials availability}
All data needed to evaluate the conclusions in the paper are present in the paper and/or the Supplementary Materials. All the code used to produce the results can be downloaded from \url{https://github.com/darioizzo/geodesynets}. Additional data related to this paper is released at \url{https://zenodo.org/record/4749715#.YJrR6OhfiUk}

\end{document}